\documentclass[prb,twocolumn, superscriptaddress]{revtex4}
\usepackage{epsfig}
\usepackage{graphicx}
\usepackage{dcolumn}
\usepackage{bm}
\usepackage[english]{babel}
\usepackage{amsmath, amssymb}
\usepackage{pdfpages}

\begin{document}
	
\title{Electrocrystallization of Supercooled Water Confined by Graphene Walls}

\author{Ramil M. Khusnutdinoff} \email{khrm@mail.ru}
\author{Anatolii V. Mokshin} \email{anatolii.mokshin@kpfu.ru}
\affiliation{Kazan (Volga region) Federal University, 420008 Kazan, Russia}
\affiliation{Udmurt Federal Research Center of the Ural Branch of the Russian Academy of Sciences, 426068 Izhevsk, Russia}
	
\begin{abstract}
Any structural transformation of water is sensitive to an external electric field, since water molecules have dipole moments. We study influence of external uniform electric field on crystallization of supercooled water enclosed between two graphene planes.  Crystallization of such the system is caused by ordinary relaxation of the metastable phase into an ordered (crystalline) phase and by dipole alignment induced by an applied electric field. We found that this system at the temperature $T=268$~K, where water has the density $0.94$~g/cm$^3$ and the applied electric field is of the magnitude $E=0.5~\textrm{V/\AA}$, crystallizes into the cubic ice with few the defects, and the crystallization proceeds over the time scale $\sim 5.0$~ns. The obtained results can be directly used to develop the methods to drive by water crystallization.
\end{abstract}

\maketitle

\section{Introduction}

Understanding the mechanisms of phase transitions in confined water is one of open problems in condensed matter physics, which stimulates performance of numerous experimental, numerical and theoretical studies \cite{Mishima1998,Debenedetti2003,Galenko2019,Galenko2019b,Chakraborty2017,Gao2018,Khusnutdinoff2011,Khusnutdinoff2012}.
So, investigations of the properties of water in a confinement attract the special attention of scientists because of its importance for understanding many biological and geological processes, such as the microscopic processes in membranes and cells; transport processes in pores; capillary phenomena; dynamic processes in the depths of satellites of some planets of the Solar system~\cite{Skripov/Faizullin,Han2010}.
Various external fields such as the external pressure~\cite{Bai2012, Murray2005}, ultrasound~\cite{Debenedetti1996}, electric~\cite{Svishchev1996, Zangi2004, Yan2011} and magnetic~\cite{Zhang2010} fields can transform the hydrogen bond network  and can induce specific ordered phases in water under confined conditions.
Crystallization of  water in a confinement is of a special interest, because it occurs in a lot of various natural and technological processes.
As an example, global natural and climatic changes are directly associated with the processes of electrocrystallization of water droplets in the Earth's atmosphere \cite{Murray2005,Whalley1981,Riikonen2000}.

Molecular dynamics simulations is a useful tool to address the issues related to phase transitions (in particular, crystallization) at the atomistic/molecular level. The simulation results for the crystallizing simple monatomic liquids and amorphous solids complement the experimental findings and provide an excellent basis to develop the microscopic theoretical models of crystal nucleation, crystal growth and overall crystallization~\cite{Mokshin2008,Sterkhova2014,Mokshin2009}. At present, numerical simulation methods are actively used to study the phase transitions in complex molecular systems. The main goal of this work is to clarify the role of a uniform electric field in crystallization of supercooled water, that corresponds to the case of the so-called \textit{electrocrystallization}. We consider the specific case when the crystallizing system represents an extremely thin layer of water enclosed between two graphene planes (see Fig.~\ref{Fig01}).

\section{The considered system and simulation details}

\begin{figure}
	\centering
	\includegraphics[width=1.0\linewidth]{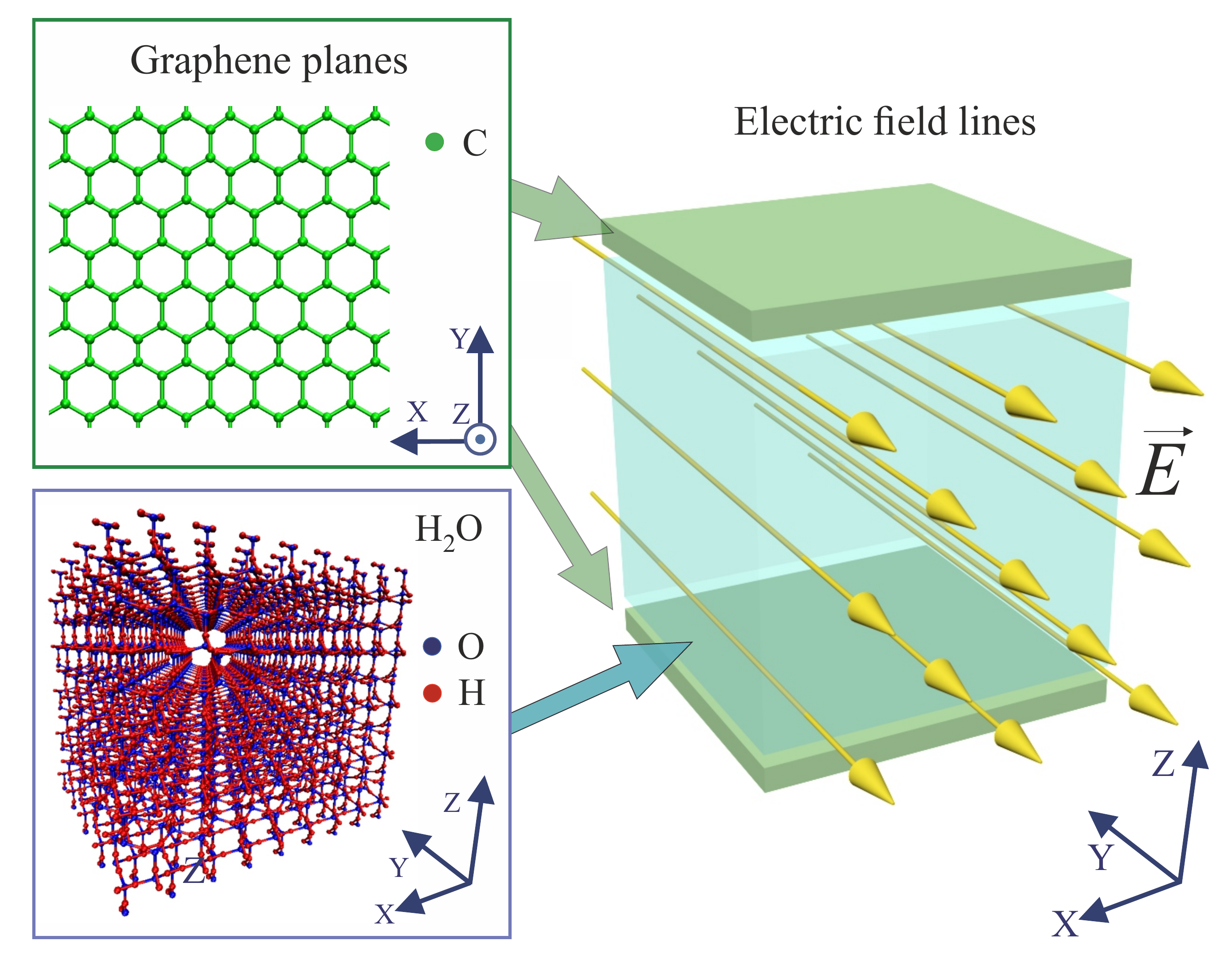}
	\caption{Schematic plot of the system under study: supercooled water is enclosed between two parallel graphene sheets, and  an external homogeneous electric field is applied to the system in the lateral direction (Y-axis).}
	\label{Fig01}
\end{figure}
The simulation cell is a sandwich consisting of water and two parallel graphene planes located on opposite sides of this cell along the $Z$-direction. The distance between the graphene planes is equal to $d=40.38$~\AA. The electric field of the fixed magnitude $E=|\vec{E}|=0.5~\textrm{V/\AA}$ is applied along $Y$-direction. This cell is shown schematically in Fig.~\ref{Fig01}. We note that the electric field with $E=0.5~\textrm{V/\AA}$ is strong enough, but it is comparable in magnitude to that experienced by water molecules near the surfaces of biopolymers \cite{Drost-Hansen} and within the cracks of amino-acid crystals \cite{Gavish1992}. Periodic boundary conditions are applied only along the $X$ and $Y$ directions.

This simulation cell contains $1530$ water molecules and $1152$ carbon atoms. The water molecules interact through the Tip4p/Ice potential \cite{Abascal2005}, which reproduces properly the equilibrium phase diagram of water~\cite{Vega2011}. The intramolecular bonds and angles are constrained by conditions according to the SHAKE-algorithm~\cite{Ryckaert1977}. Interactions between water molecules and carbon atoms is given by the Lennard-Jones potential \cite{Gordillo2010}, where the parameters of the interaction -- the effective size and energy - are given by the Lorentz-Berthelot mixing rule. To take into account the long-range Coulomb interactions between the partial charges, we use the PPPM method with the cutoff radius $r_c=\textrm{13~\AA}$~\cite{Rajagopal1994}. The locations of carbon atoms are fixed.

Molecular dynamics simulations were performed in the NVT ensemble for the system with the temperature $T=268$~K and the mass density $0.94$~g/cm$^3$. The Nose-Hoover thermostat with a damping constant of $1.0$~ps is applied to keep the isothermal conditions \cite{Nose1984,Hoover1985}. The sample with the supercooled water at the temperature $T=268$~K was generated by fast cooling with the rate $\gamma=10^{12}$~K/s from the high-temperature liquid state at the temperature $T=350$~K.

\section{Results and discussion}

As was shown before \cite{Khusnutdinoff2013}, an external electric field can promote the structural ordering in the similar system with the polar molecules. When we apply the electric field of various fixed magnitudes (from $E=0$ to $E=1.0~\textrm{V/\AA}$) along the lateral direction of the sandwich-system as shown in Fig.~\ref{Fig01}, we detect the structural changes induced by the field, and these structural changes are accompanied by appearance of a non-zero polarization vector. In particular, the crystallization induced by the field is  directly evidenced by the features of the radial and angular distribution functions evaluated for the oxygen atoms of the water molecules. The peaks in these functions become more pronounced, that is typical for the structural ordering.
\begin{figure}
	\centering
	\includegraphics[width=2.2\linewidth]{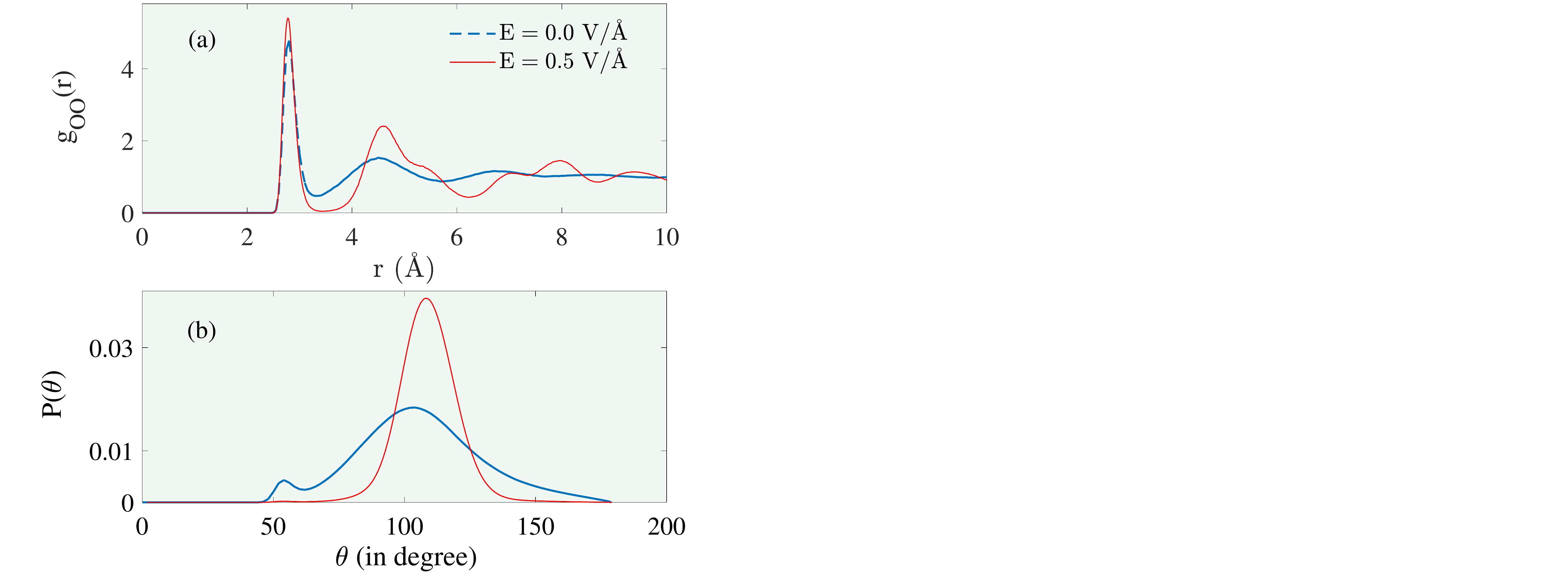}
	\caption{Oxygen-oxygen radial (a) and angular (b) distribution functions for the confined supercooled water at the temperature $T=268$~K and density of $0.94$~g/cm$^3$ without electric field ($E=0$) and with electric field of the magnitude $E=\textrm{0.5~V/\AA}$.}
	\label{Fig02}
\end{figure}

In Fig.~\ref{Fig02} we present the radial and angular distribution functions, $g(r)$ and $P(\theta)$, for the system, when it is at rest ($E=0$) and when electric field with $E=\textrm{0.5~V/\AA}$ is applied. For the case of $E=\textrm{0.5~V/\AA}$, the functions shown in this figure correspond to the state of the system when it has completely reacted to the imposed electric field. As seen for this case, long-range correlations in the radial distribution function $g(r)$ extend up to the distance $\sim 10$~\AA, that is signature of  the translational order. The  angular distribution function $P(\theta)$ takes the form of the Gaussian function located at $\theta \simeq 108^{o}$, that is typical for the local structures with the tetrahedral symmetry.

\begin{figure}
	\centering
	\includegraphics[width=1.7\linewidth]{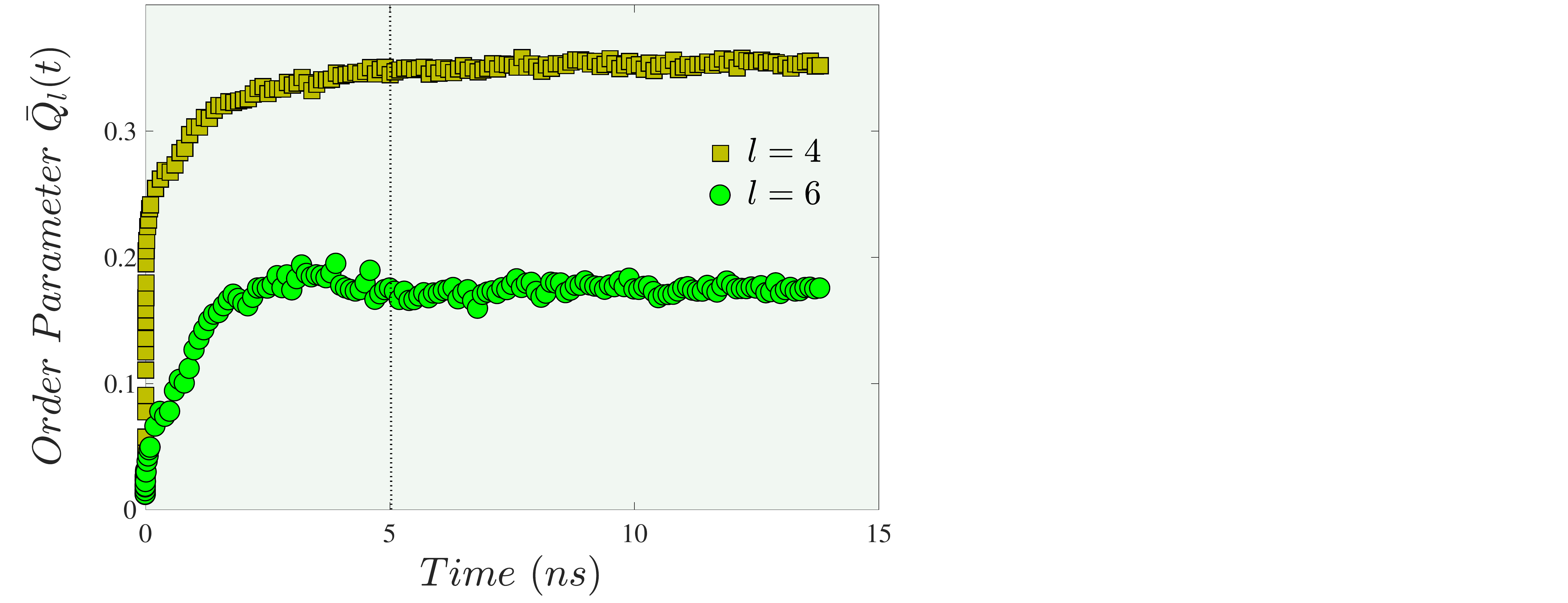}
	\caption{Time dependence of the global orientational order parameters $\bar{Q}_4$ and $\bar{Q}_6$ evaluated for the considered system under applied electric field (see discussion in the text).}
	\label{Fig03}
\end{figure}

The local structure of the system can be analyzed in detail using the cluster analysis based on computation of the orientational order parameters~\cite{Steinhardt1983,Lechner2008}.
As was shown in Ref. \cite{Brukhno2008}, the local order parameters of the standard Steinhardt-Nelson-Ronchetti's scheme~\cite{Steinhardt1983} are not sensitive to detect the proton-ordered hexagonal (Ih) and cubic (Ic) ices.
For the reason, one use the following set of the orientational order parameters~\cite{Lechner2008}, namely, the adopted local order parameter
\begin{equation}
\bar{q}_{l}(i)=\left(\frac{4\pi}{2l+1}\sum_{m=-l}^{l}\left|\bar{q}_{lm}(i)\right|^{2}\right)^{1/2}
\label{eq_q}
\end{equation}
and the adopted global order parameter
\begin{equation}
\bar{Q}_{l}=\left(\frac{4\pi}{2l+1}\sum_{m=-l}^{l}\left|\frac{\sum_{i=1}^{N}n_{b}^{(i)}\bar{q}_{lm}(i)}{\sum_{i=1}^{N}n_{b}^{(i)}}\right|^{2}\right)^{1/2},
\label{eq_Q6_bop}
\end{equation}
where
\begin{equation}
\bar{q}_{lm}(i)=\frac{1}{1+n_{b}^{(i)}}\sum_{j=1}^{n_{b}^{(i)}}\frac{1}{n_{b}^{(j)}}\sum_{k=1}^{n_{b}^{(j)}}Y_{lm}(\theta_{jk},\varphi_{jk})
\nonumber
\end{equation}
and
\[
l=2,\,4,\, 6,\ \ldots .
\]
In Eqs.~(\ref{eq_q}) and (\ref{eq_Q6_bop}), the quantity $n_{b}^{(i)}$ is the number of neighbors of an $i$th molecule; $Y_{lm}(\theta_{ij},\varphi_{ij})$ are the spherical harmonics;
$\theta_{ij}$ and $\varphi_{ij}$ are the polar and azimuthal angles, respectively.
The local order parameter $\bar{q}_{l}(i)$ is evaluated for each particle of the system (of an instantaneous configuration). Then, the distribution of the parameter $P(\bar{q}_{l}(i))$ must have the specific forms for different crystalline phases and a disordered phase.
The order parameters $\bar{Q}_l$ with $l = 4, 6, 8,...$ take unique values for the specific crystalline structures, and, therefore, values of these parameters allow one on the basis of the known coordinates of all the molecules to identify the symmetry of the arrangement of these molecules~\cite{Steinhardt1983}.  For example,  the parameter $\bar{Q}_4$ takes values $0.259$ and $0.506$ for the perfect proton-ordered hexagonal and cubic ice of low density, respectively. For a fully disordered system, one has $\bar{Q}_4 \to 0$. Further, the parameter $\bar{Q}_6$ is not sensitive to recognize whether there is hexagonal or cubic ices, but it takes different values for the crystalline ices and water~\cite{Galimzyanov2019}.
\begin{figure}
	\centering
	\includegraphics[width=2.2\linewidth]{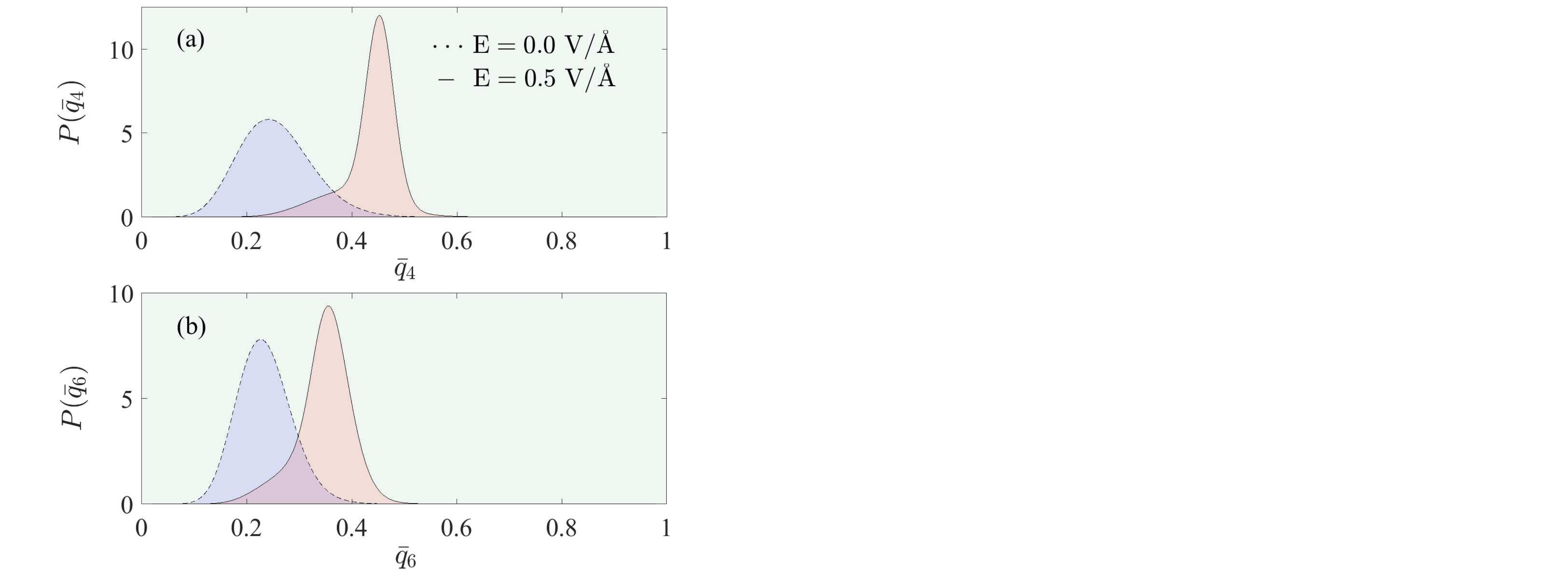}
	\caption{Distributions of the local order parameters $\bar{q}_4$ and $\bar{q}_6$ evaluated for the system at rest and for the system with applied electric field (see discussion in the text). Panel (a) corresponds to the distribution $P(\bar{q}_4)$, and panel (b) shows results for the distribution $P(\bar{q}_6)$.}
	\label{Fig04}
\end{figure}

Let us now present results for the system with applied electric field $E=0.5~\textrm{V/\AA}$. In Fig.~\ref{Fig03}, we present time dependence of the global order parameters $\bar{Q}_4(t)$ and $\bar{Q}_6(t)$. As seen, these parameters demonstrate sharp increase within a nanosecond scale and the subsequent smooth saturation to values $\bar{Q}_4=0.35$  and $\bar{Q}_6=0.175$. Thus, the quasi-equili\-bri\-um state is achieved over the time scale $\tau\approx 5$~ns. It is remarkable that value of the parameter $\bar{Q}_4(t)$ for this quasi-equilibrium state is larger than $\bar{Q}_4^{(Ih)}=0.259$, but lesser than $\bar{Q}_4^{(Ic)}=0.506$. This can indicate that the obtained phase is an ordered one. We suppose that there are two alternatives: this phase can represent a cubic ice with defects or a mix of cubic ice and hexagonal ice.

\begin{figure}
	\centering
	\includegraphics[width=1.0\linewidth]{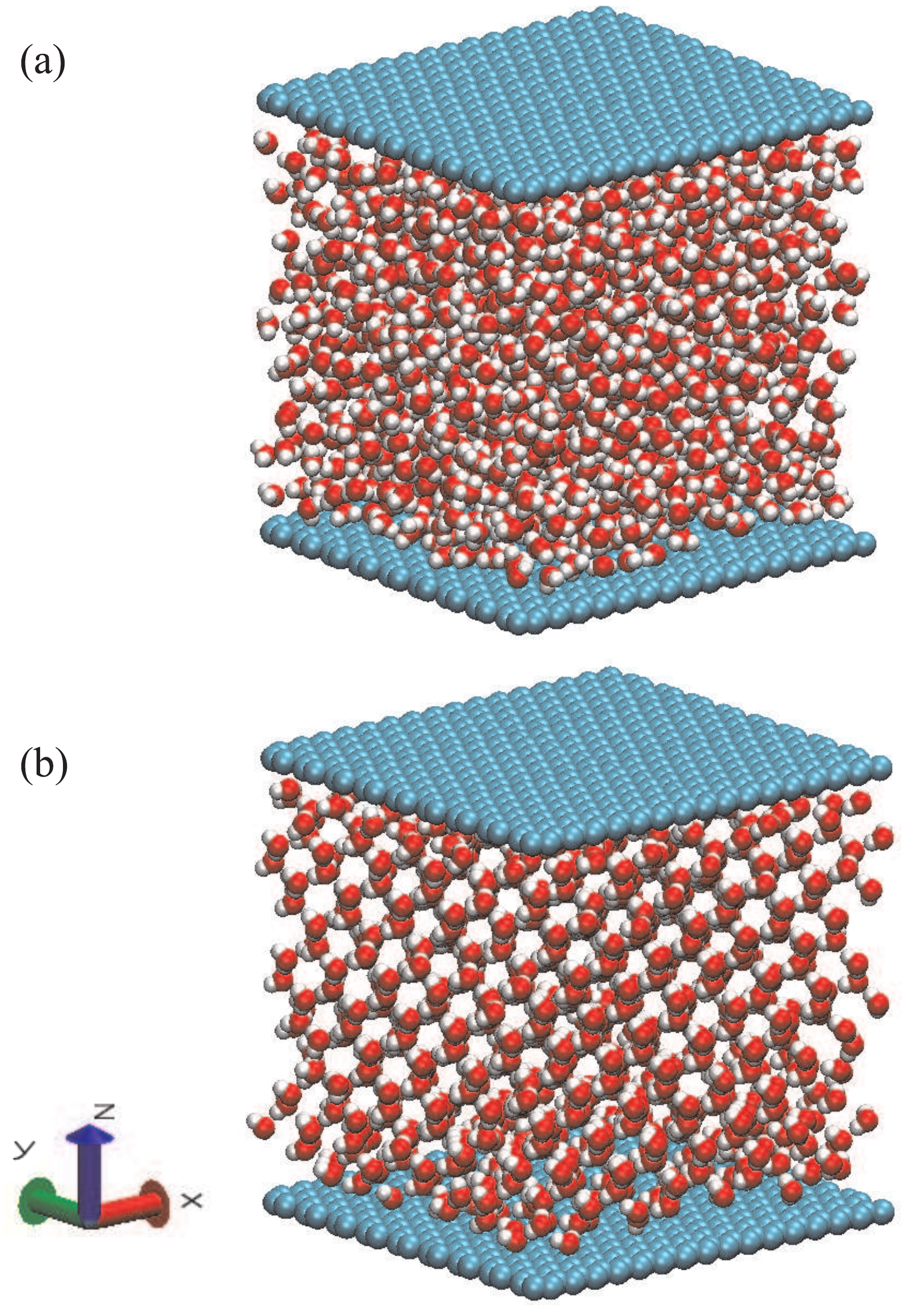}
	\caption{Instantaneous snapshots of the simulation cell with the supercooled water (a) and with the quasi-equilibrium state obtained with the applied external electric field (b).}
	\label{Fig05}
\end{figure}
In Fig.~\ref{Fig04}, we present the distributions $P(\bar{q}_4)$ and $P(\bar{q}_6)$ of local order parameters for the system with applied electric field for the time moment $t=14$\,ns, i.e. when a quasi-equilibrium state is achieved. As seen, both the distributions are located at values of the order parameters larger then for the case of a disordered system, that is typical for an ordered phase. On the other hand, both the distributions are not symmetric and have shoulders at low values of the order parameters corresponding to a disordered system.

Not large size of the simulation cell allows one to perform direct visual inspection of the structure obtained (Fig.~~\ref{Fig05}). The hexagonal channels typical for the cubic ice in the crystallographic [111]-direction are clearly seen in snapshot given in Fig.~\ref{Fig05}(b). Thus, due to the applied field, the system form high-polarized cubic ice. This is remarkable for the following reason. According to the equilibrium phase diagram of bulk water, for the given density/temperature thermodynamic state, the hexagonal ice structure is the most stable.  The phase of the cubic ice appears only for the low temperatures $T \in [123; 153]$~K at ambient pressure. In this regard we note that both the crystalline phases - Ic and Ih - are similar and have an almost perfect tetrahedral geometry~\cite{Wang2018}. Moreover, the basal (001)-plane of ice-Ih and the (111)-plane of ice-Ic are identical.
Under the certain conditions, the cubic ice recrystallizes into the hexagonal ice, that was experimentally observed (see Ref.~\cite{Murray2005}).
Finally, as was mentioned in Ref. \cite{Johari2005},  the cubic ice is actually the more stable crystalline phase for water confined in pores or bewteen the thin films of the linear size of the confinement $\Delta z\leq 100~\textrm{\AA}$. This is in agreement with the results of this study.

\section{Concluding remarks}
The electromagnetic field is one of the factors by means of which the structural and dynamical properties of water can be modified~\cite{Guidelli1992}. In this study we have shown that the electric homogeneous field has direct impact on water crystallization. Namely, this field promotes the structural ordering of water confined between two graphene layers and favors to formation of the high-polarized phase of the cubic ice. The given results can be useful for development of the methods to drive by the crystallization of water in confined geometry that is of great importance in biology, medicine, physiology and geology.

\section{Acknowledgements}
We thank  Bulat N. Galimzyanov for his help to prepare Fig. 1.
This work is supported by the Russian Science Foundation (project No. 19-12-00022). The molecular dynamic simulations were performed by using the computational cluster of Kazan Federal University and the computational facilities  of Joint Supercomputer Center of RAS.

\bibliographystyle{unsrt}

\end{document}